\documentclass[aps,prl,twocolumn,graphicx,amsbsy,amssymb,showpacs]{revtex4}

\usepackage{graphicx}
\usepackage{amsmath}

\begin{document}
\title{ Origin of long-ranged attraction between like-charged particles at water-air interface}
\author{Yi Zhou and Tai-Kai Ng}
\affiliation{
Department of Physics,
Hong Kong University of Science and Technology,
Clear Water Bay Road,
Kowloon, Hong Kong
}
\date{ \today }
\begin{abstract}
  The nature of electrostatic interaction between like-charged particles at water-air interface is analyzed in
 this paper. We show that long-ranged electrostatic dipolar attraction between these objects generally exists.
 Our result provides a natural mechanism to explain the experimental
 observations of attraction between like-charge species trapped at water-air interface. We speculate that a similar
 (but presumably weaker) mechanism also exists for particles near water-solid interface.
\end{abstract}


\maketitle

   When charged colloidal particles are dispersed into water, a fraction of the ionic functional groups on the
 surface dissociate. The net charge remaining on the particle surface is screened by the cloud of counter-ions in
 water, resulting in a screened Coulomb potential between particles\cite{sc}
 \[
 U(r)={q^2\over\epsilon_w(1+a/\lambda_D)^2}{e^{-(r-2a)/\lambda_D}\over r},
 \]
 where $r$ is the sphere's center-to-center separation, $q$ is the effective charge carried by each particle with
 radius $a(<<\lambda_D)$, $\epsilon_w$ is the dielectric constant of water and $\lambda_D$ is the Debye-H\"{u}ckel
 screening length. This screened Coulomb repulsion together with short-ranged Van der Waals attraction usually
 determines the phase diagram of charged colloidal systems\cite{sc,col}.

  In recent years, however, there has been increasing number of experiments reporting that under special
  circumstances, attractive interactions rather than the screened-Coulomb repulsion exist between like-charged
  species ranging from simple colloidal particles\cite{e1,e2,e3,e4,e5} to complex cytoskeletal filamentous
  actin\cite{e6} and DNA\cite{e7}. Because of its fundamental interest and important implications in colloid
  science and biology, the problem has been under intensive theoretical study\cite{t1,t2,t3,t4,t5,t6,t7} for many
  years.

\begin{figure}
   \includegraphics[width=6.0cm, angle=0]{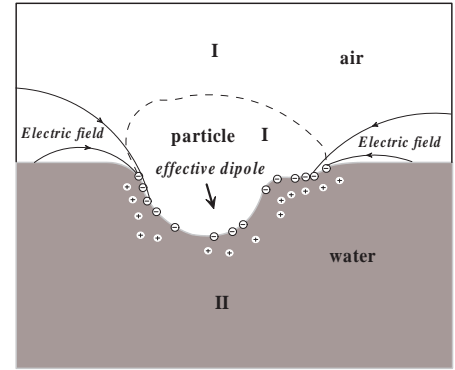}
   \caption{\label{Fig.1} A charged particle trapped in a water-air interface.}
   \end{figure}

   In this paper, we shall study electrostatic interaction between charged particles trapped at water-air
  interface. It was discovered that colloidal particles that repel each other in deep water may attract each
  other when they are close to water-air\cite{ls1,ls2,ls3,ls4} or water-solid\cite{e1,e2,e3,e4,e5} interfaces.
  Moreover, whereas like-charge attraction between particles deep inside water usually occurs at distance scale
  $r\leq\lambda_D$\cite{e6,e7,ls3}, attraction between colloidal particles near interface is found to appear
  at length scale $r>>\lambda_D$\cite{chen}. The long-range-ness of the interaction suggests that the origin of
  attraction is probably independent of details like nature of the screening charge or chemical environment
  which largely determines the interaction at length scale $\leq\lambda_D$. Indeed, it has been
  proposed\cite{dipole1,dipole2} that because one of the phases forming the interface (air or solid) is made of
  a non-polar substance that cannot sustain charge, the counter-ions in water are distributed asymmetrically
  around the particle, producing an effective dipole moment for each particle (Fig. \ref{Fig.1}). For a spherical
  particle with uniform charge distribution on the particle surface, the resulting dipole moment points downward
  perpendicular to the interface, leading to long-ranged dipolar ($\sim 1/r^3$) repulsion between particles.
  Following this line of argument, it was further suggested that charge inhomogeneity on the surface may lead to
  formation of an effective dipole moment {\em not} perpendicular to the interface, and may lead to long-range
  dipolar attraction between particles\cite{chen}.

    Whereas this argument is physically suggestive, it has not been proven with rigor that the mechanism works.
  An electric dipole inside water is completely screened as monopole charges, and the interaction between dipoles
  becomes short-ranged. It is not clear that the $1/r^3$ long-range interaction will survive for an effective
  electric dipole trapped at the surface of water, or whether strong correction to the dipolar interaction exists.
  The purpose of our paper is to clarify this problem and show that the dipole picture is valid in the limit when
  $\lambda_D$ is smaller than any other length scales in the system and dipolar attraction between particles
  located at water-air interface appears under very general conditions that are independent of microscopic details
  of the system.

    We start by considering a particle of arbitrary
   shape trapped at a water-air interface (Fig. \ref{Fig.1}). Charges from the particle surface in contact with water
   dissociate leaving a surface layer of charge $\sigma(\vec{x})$ at the particle-water interface. The dielectric
   constant of the particle is $\epsilon_p$. Therefore the electric potential arising from
   nonzero $\sigma(\vec{x})$ is obtained by solving the Poisson equation
	 \begin{subequations}
   \label{p1}
   \begin{equation}
   \label{pos}
   \nabla^2\Phi^{(I)}=4\pi\nabla\cdot\vec{P},
   \end{equation}
   at region I (air+particle) where
   \begin{equation}
   \label{pol}
    \vec{P}={\epsilon_p-1\over4\pi}\vec{E}^{(I)}={1-\epsilon_p\over4\pi}\nabla\Phi^{(I)},
  \end{equation}
    is the electric polarization inside the particle and $\vec{E}^{(i)}=-\nabla\Phi^{(i)} (i=I,II)$ is the electric field
  at region $i$. The electric potential at region II (water) is obtained by solving the linear Poisson-Boltzmann equation
   \begin{equation}
   \label{pb}
    (-\nabla^2+\lambda_D^{-2})\Phi^{(II)}=0,
   \end{equation}
   subject to the boundary condition
   \begin{equation}
   \label{bc}
   [\epsilon_p\vec{E}^{(I)}(\vec{x})-\epsilon_w\vec{E}^{(II)}(\vec{x})]\cdot\hat{n}(\vec{x})=4\pi\sigma(\vec{x}),
   \end{equation}
   \end{subequations}
  where $\vec{x}$ runs over the surface $S$ separating region I and II and $\hat{n}(\vec{x})$ is a unit vector
  normal to $S$ (hereafter we will specify $\hat{n}$ pointing from regions II to I). Electric polarization is
  introduced explicitly at region I for convenience as we shall see later. It is not introduced in water explicitly.

    Although exact solution to the above electrostatic problem is not available in general, the nature of the
  solution can be seen by using the Green's Theorem\cite{jackson} which states that the potential
  $\Phi^{(I)}$ at region I can be expressed in the form
  \begin{eqnarray}
  \label{g1}
  \Phi^{(I)}(\vec{x})&=&\int_Vd^3x'{\nabla'\cdot\vec{P}(\vec{x}')\over R}-{1\over4\pi}\oint_S\left[{1\over R}
  \frac{\partial\Phi^{(I)}(\vec{x}')}{\partial n'}\right.\nonumber\\
  &&\left.-\Phi^{(I)}(\vec{x}')\frac{\partial}{\partial n'}({1\over R})\right]da'
  \end{eqnarray}
   where $R=|\vec{x}-\vec{x}'|$ and $da'$ runs over the surface separating the two regions. The surface integral
   can be interpreted as the potential due to a surface charge density $\sigma_I=-{1\over4\pi}
   \epsilon_p\frac{\partial\Phi^{(I)}}{\partial n'}$ and dipole density $\vec{p}_I={1\over4\pi}\Phi^{(I)}\hat{n'}$\cite{jackson}.

   A similar theorem exists at region II with the Green's function $1/R$ for the Poisson equation in free space
   replaced by the corresponding Green's function $(e^{-R/\lambda_D})/R$ for the Poisson-Boltzmann equation, i.e.
   \begin{eqnarray}
   \label{g2}
   \Phi^{(II)}(\vec{x})={1\over4\pi}\oint_S\left[{e^{-R/\lambda_D}\over R}\frac{\partial\Phi^{(II)}(\vec{x}')}{\partial n'}\right.\nonumber \\
   \left.-\Phi^{(II)}(\vec{x}')\frac{\partial}{\partial n'}({e^{-R/\lambda_D}\over R})\right]da'.
   \end{eqnarray}
    Notice there is no "space charge" at region II. We note that the Green's theorem does not provide a full
    solution to the boundary-value problem, but only an
    integral statement since arbitrary specification of both $\Phi$ and $\frac{\partial\Phi}{\partial n'}$ is an
    over-specification of the problem. Nevertheless, the theorem dictates the form of the solution. The resulting
    potential of any charge distribution $\sigma(\vec{x})$ on the surface separating the two regions can be viewed
    as coming from effective charge and dipole (normal to the surface) distributions on the same surface plus
    induced electric polarization inside the particles in {\em free space}. Interaction between particles are
    determined if we can find these effective charge and dipole distributions.

      Eq.\ (\ref{pb}) can be solved easily in the limit when $\lambda_D$ is much less than the two other length
    scales in the problem, the length scales where $\sigma(\vec{x})$ varies and the surface curvature of the
    interface. In this case, we can assume that the surface $S$ is locally flat and $\sigma(\vec{x})$
    is locally constant. The solution to Eq.\ (\ref{pb}) at region II not far from the interface is
    approximately
    \[
    \Phi^{(II)}(x',y',z'))\sim\Phi(x',y',0)e^{-|z'|/\lambda_D},  \]
    where $(x',y',0)$ denotes a point on the interface $S$ and $\vec{z}'=z\hat{n}'$ is a vector pointing away
    from the interface at point $(x',y')$. Correspondingly,
    \[
    \frac{\partial\Phi^{(II)}}{\partial n'}\sim-{1\over\lambda_D}\Phi^{(II)},  \]
   on the interface. Putting this back into Eq.\ (\ref{g1}), integrating by part and making use of the boundary
   condition\ (\ref{bc}), we obtain at region I
   \begin{eqnarray*}
   \Phi^{(I)}(\vec{x})&=-\int_Vd^3x'\vec{P}(\vec{x}')\cdot\nabla'({1\over R})\nonumber
   +{1\over4\pi}\oint_S\left\{\Phi_S(\vec{x}')\frac{\partial}{\partial n'}({1\over R})\right.\nonumber\\
   &\left.+{1\over R}[4\pi\sigma(\vec{x}')-{\epsilon_w\over\lambda_D}\Phi_S(\vec{x}')]\right\}da',
   \end{eqnarray*}
   where $\Phi^{(I)}=\Phi^{(II)}=\Phi_S$ on the interface. $\Phi_S$ can be determined easily
   if we impose the physical requirement that $\Phi^{(I)}(\vec{x})\rightarrow0$ everywhere in region I in the
   metallic limit $\lambda_D\rightarrow0$\cite{jackson}. In this limit $\Phi_S(\vec{x}')=
   4\pi\lambda_D\sigma(\vec{x}')/\epsilon_w$ and
   \begin{equation}
   \label{g4}
   \Phi^{(I)}(\vec{x})\rightarrow-\int_Vd^3x'\vec{P}(\vec{x}')\cdot\nabla'({1\over R})+
   \oint_S{\sigma(\vec{x}')\lambda_D\over\epsilon_w}\frac{\partial}{\partial n'}({1\over R})da'.
   \end{equation}
   corresponding to the potential from a volume electric polarization $\vec{P}$ and surface dipole density
   $\vec{p}(\vec{x}')=\lambda_D\sigma(\vec{x}')\hat{n}(\vec{x}')/\epsilon_w$. More generally,
   a systematic expansion of $\Phi_S$ in powers of $\lambda_D$ can be set up, and Eq.\ (\ref{g4}) gives the
   leading order result. Notice that Eq.\ (\ref{g4}) should be solved together with Eq.\ (\ref{pol}) to determine
   $\vec{P}$ and $\Phi^{(I)}$. Although analytical solution cannot be obtained in general, the dipolar nature of
   the electric potential of charge particles at the air-water interface is apparent.

     The interaction between two particles with surface charge distributions $\sigma_1(\vec{x}')$ and
   $\sigma_2(\vec{x}')$ is given by the cross-term in the total electrostatic energy,
   $W_{12}={1\over4\pi}\int^{(I)}d^3x\epsilon_p\vec{E}_1(\vec{x})\cdot\vec{E}_2(\vec{x})+
   {\epsilon_w\over4\pi}\int^{(II)}d^3x\left[\vec{E}_1(\vec{x})\cdot\vec{E}_2(\vec{x})
   +\lambda_D^{-2}\Phi_1(\vec{x})\Phi_2(\vec{x})\right]$, where the first term is the contribution from region I
   and the second term from region II. The first term dominates in the limit
   $\lambda_D\rightarrow0$ and
    \begin{equation}
    \label{int}
    W_{12}\sim\int_{V_1}dv_1\int_{V_2}dv_2(\vec{P}_1\cdot\nabla_1)(\vec{P}_2\cdot\nabla_2)G^{(I)}(\vec{x}_1,\vec{x}_2),
    \end{equation}
    where $\int_{V}dv\vec{P}..$ includes both the volume and surface contributions to the electric dipole and
    \begin{equation}
    \label{int1}
    G^{(I)}(\vec{x}_1,\vec{x}_2)={1\over|\vec{x}_1-\vec{x}_2|}+{1\over8\pi}\oint_{S}(\hat{n}(\vec{x})\cdot\nabla)
    {1\over|\vec{x}-\vec{x}_1||\vec{x}-\vec{x}_2|}da.
    \end{equation}

     The first term in Eq.\ (\ref{int1}) gives rise to usually dipole-dipole interaction in free space whereas the
    second term is a correction term coming from the water-air interface. It is not difficult to show that
    $\hat{n}(\vec{x})\cdot\nabla(|\vec{x}-\vec{x}_1||\vec{x}-\vec{x}_2|)^{-1}=0$ for flat surface and the second term
    is generally smaller than the first term. The naive dipole picture\cite{dipole1,dipole2,chen} for
    interaction between colloids on water-air interface is recovered if we neglect the volume contribution
    $\vec{P}$ and the surface correction term in $G(\vec{x}_1,\vec{x}_2)$.

      We next discuss corrections from finite value of $\lambda_D$. These corrections are generally small when the
    particles are far away from each other but may become large and dominant when they are within distance
    $\sim\lambda_D$. The most important contribution is perhaps the region II contribution which is given by a
    formula similar to $W_{12}$,
      \begin{eqnarray}
      \label{int2}
      W_{12}^{(II)}&=&\epsilon_w\oint_{S_1}da_1\oint_{S_2}
      da_2({\sigma_1\over\epsilon_w}+\vec{p}_1\cdot\nabla_1)\nonumber \\
      &&\times({\sigma_2\over\epsilon_w}+\vec{p}_2\cdot\nabla_2)G^{(II)}(\vec{x}_1,\vec{x}_2),
      \end{eqnarray}
      where $\sigma_i$ is the surface charge density for particle $i$ and $\vec{p}_i$ is the
      corresponding surface dipole moment as discussed before. Notice that $\vec{P}$ does not contribute in
      this region. $G^{II}$ is the Green's function for the Poisson-Boltzmann equation,
      \begin{eqnarray}
      \label{G2}
      G^{(II)}(\vec{x}_1,\vec{x}_2)&=&{e^{|\vec{x}_1-\vec{x}_2|/\lambda_D}\over|\vec{x}_1-\vec{x}_2|}
      -{1\over8\pi}\oint_{S}(\hat{n}(\vec{x})\cdot\nabla)\nonumber\\
      &&\times{e^{|\vec{x}_1-\vec{x}_2|/\lambda_D}\over|\vec{x}-\vec{x}_1||\vec{x}-\vec{x}_2|}da.
      \end{eqnarray}
       The surface charges $\sigma_i's$ with the first term in $G^{(II)}$ give rise to the celebrated screened
      Coulomb interaction between particles. The second term in $G^{(II)}$ is a corresponding interface correction.

   \begin{figure}[htpb]
   \label{Fig.2}
   \vspace{-0.5cm}\includegraphics[width=6.0cm]{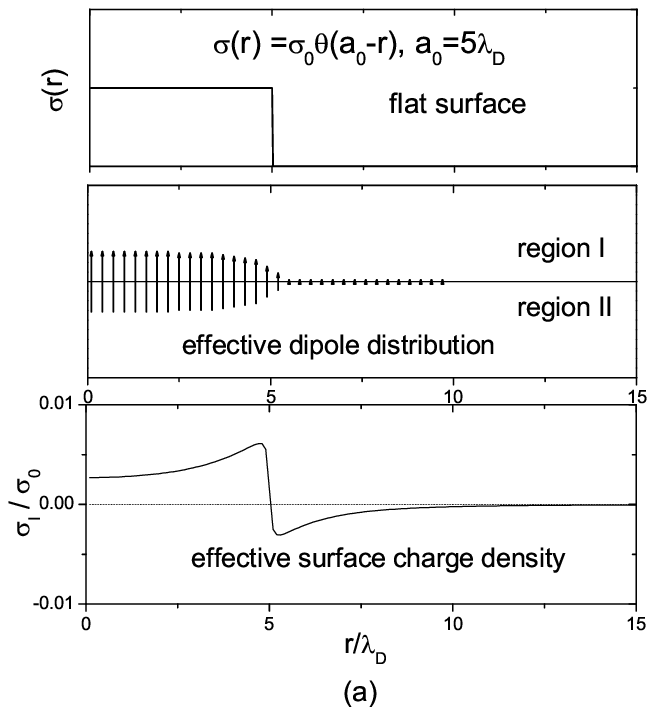}\vspace{-0.5cm}
   \includegraphics[width=6.0cm]{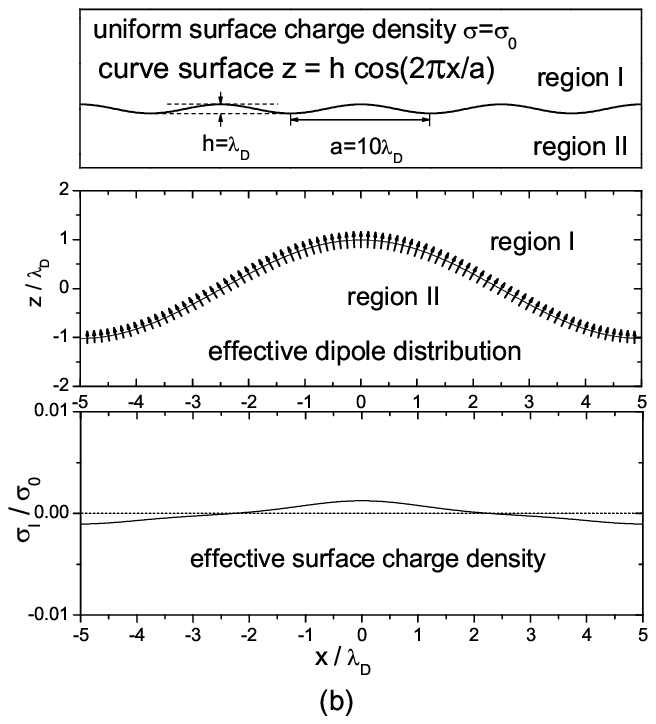}\vspace{-0.5cm}
   \caption{\label{Fig.2} Effective dipole distribution with dipole density $\vec{p}_I={1\over4\pi}\Phi_S\hat{n}$ and
   effective surface charge with charge density $\sigma_I=-{1\over4\pi}\epsilon_p\frac{\partial\Phi^{(I)}}{\partial n}$ for
   two different situations.
   (a) A non-uniform charge distribution $\sigma(\vec{x})=\sigma_0\theta(a_0-r)$ on a flat surface,
   all the charges appear inside a disk of radius $a_0$, where $r$ is the radial coordinate on the surface
   and $a_0=5\lambda_D$. (b) An uniform charge distribution $\sigma(\vec{x})=\sigma_0$ on a curve
   surface $z=h\cos(2\pi x/a)$, $h=\lambda_D$ and $a=10\lambda_D$.}
   \end{figure}
   
      Further corrections to effective dipole and surface charge distribution exist.
    These corrections reflect finite surface curvature or spatial dependence of
    $\sigma(\vec{x})$ and are important when the length scale is comparable to the screening
    length $\lambda_D$. In this case the counter-ions in the fluid cannot adjust themselves to these rapid changes
    fully, resulting in formation of small patches of local charges or correction to magnitude of dipole moment
    at length scale $\sim\lambda_D$. Notice that because of the long-rangeness of the Coulomb potential, non-local
    corrections to $\Phi_S$ also arises and are important when surface curvature is nonzero. To
    illustrate we show in Fig. \ref{Fig.2} numerical solutions of $\Phi^{(I)}$ and $\partial\Phi^{(I)}/\partial n$ for
    two different situations: (i) a non-uniform charge distribution $\sigma(\vec{x})=\sigma_0\theta(a_0-r)$ on a flat surface,
    which is a two dimensional disk (Fig. \ref{Fig.2}(a)), and (ii) an uniform charge distribution $\sigma(\vec{x})=\sigma_0$
    on a curve surface $S$ given by $z(x,y)=h\cos(2\pi x/a)$, which is stripe like (Fig. \ref{Fig.2}(b)).
    Water occupies the lower half of the figure (region II). We see that both surface curvature and non-uniform
    surface charge distribution give rise to corrections to $\Phi_S$ and $\partial \Phi^{(I)}/\partial n$.
    For the first situation (Fig. \ref{Fig.2}(a)), the correction to effective dipole is $\delta p_I\sim-
    {\lambda^2_D\over a_0}{\epsilon_p\over\epsilon^2_w}\sigma_0$ and the correction to effective
    surface charge $\delta\sigma_I\sim{\lambda_D\over a_0}{\epsilon_p\over\epsilon_w}\sigma_0$.
    For the second situation (Fig. \ref{Fig.2}(b)), $\delta p_I\sim 2\pi^2
    {z\lambda_D^2\over a^2}{\sigma_0\over\epsilon_w}$
    and $\delta\sigma_I\sim 4\pi^3{ z\lambda_D^2\over a^3}{\epsilon_p\over\epsilon_w}\sigma_0$.

        Summarizing, we study in this paper the nature of electrostatic interaction between charged particle
    trapped at the water-air interface. We find that besides screened Coulomb interaction, long-range dipolar
    interaction generally exists between particles trapped at water-air interface, independent of microscopic
    details of the system. The interaction between particles can become attractive easily in the presence of
    non-uniform charge distribution on particle surface\cite{chen}, or for uniform charge distribution but when
    the shape of the particle is asymmetric (absence of inversion symmetry along one axis parallel to water
    surface). In both cases, a nonzero average {\em planar} electric dipole moment may be induced, and leads to
    dipolar attraction between particles. We note that a similar (but weaker) mechanism presumably exists for
    particles near a water-solid interface, if the distance between particle and the interface is less than the
    screening length $\lambda_D$. In this case, the electric charge on the particle may not be totally screened
    and some of the electric field can leak out to the non-polar solid medium, leading to dipolar interaction
    between particles. We shall investigate this situation in more details in a separate paper.

      Lastly, we emphasize that the dipolar interaction we studied in this paper is limited to particle trapped at,
    or close to interfaces where the interaction is {\em long-ranged} in nature. It probably cannot provide a
    full explanation for like-charge attraction between particles deep in water, where interaction occurs mainly at length
    scale $\leq\lambda_D$ and many other factors contribute. We note also that there exists other mechanism
    which may induce long-ranged interaction between particles trapped at water-air interaction, for example,
    van der Waals interaction\cite{col} and electric-field induced capillary attractions\cite{ls3,ls4}. Our work
    provides a solid justification for the existence of the electric dipolar interaction. Which mechanism dominates
    depends on actual experimental conditions which probably differ for different experiments.

  \acknowledgements
  We thank Dr. Wei Chen and Prof. Penger Tong for many helpful discussions and comments.  This work is supported by HKUGC through
  grant CA05/06.SC04.

\end{document}